\documentclass[twocolumn]{jpsj2} 
\title{Specific Heat and Superfluid Density for Possible Two Different Superconducting States in Na$_x$CoO$_2 \cdot y$H$_2$O}

\author{Masahito \textsc{Mochizuki}\thanks{E-mail address:
mochizuki@riken.jp}, H. Q. \textsc{Yuan}$^{1}$, and Masao \textsc{Ogata}$^{2}$}

\inst{RIKEN, Hirosawa, Wako, Saitama 351-0198, Japan\\
$^1$Department of Physics, University of Illinois at Urbana-Champaign, Urbana, Illinois 61801, U.S.A.\\
$^2$Department of Physics, University of Tokyo, Hongo, Bunkyo-ku, Tokyo 
113-0033, Japan}

\abst{Several thermodynamic measurements for the cobaltate superconductor, Na$_x$CoO$_2 \cdot y$H$_2$O, have so far provided results inconsistent with each other. In order to solve the discrepancies, we microscopically calculate the temperature dependences of specific heat and superfluid density for this superconductor. We show that two distinct specific-heat data from Oeschler $et$ $al.$ and Jin $et$ $al.$ are reproduced, respectively, for the extended $s$-wave state and the $p$-wave state. Two different superfluid-density data are also reproduced for each case. These support our recent proposal of possible two different pairing states in this material. In addition, we discuss the experimentally proposed large residual Sommerfeld coefficient and extremely huge effective carrier mass.}

\kword{Co-oxide superconductor, specific heat, superfluid density}

\begin{document}
\maketitle
The Co-oxide superconductor Na$_x$CoO$_2 \cdot y$H$_2$O has attracted great interest since its discovery.~\cite{Takada03} Several experiments have suggested an unconventional superconductivity due to a spin-fluctuation mechanism.~\cite{Takada05} 

A number of thermodynamic measurements have been performed to elucidate its superconducting (SC) state. 
However, the reported results on the upper critical field ($H_{c2}$), specific heat ($C(T)$) and superfluid density ($\rho_s(T)$) are rather scattered and contain a lot of discrepancies. As for $H_{c2}$, some groups reported values apparently exceeding the Pauli limit of the weak-coupling theory, while the other groups reported values comparable to, or less than the limit.~\cite{Takada05} As for $C(T)$, Jin $et$ $al.$ observed a shoulder-shaped structure accompanied by a steeper decrease at low temperatures,~\cite{Jin05} while Yang $et$ $al.$ and Oeschler $et$ $al.$ did not observe such a structure.~\cite{Yang05,Oeschler05} As for $\rho_s(T)$, Uemura $et$ $al.$ and Kanigel $et$ $al.$ reported a linear evolution with decreasing temperature from $\mu$SR measurements,~\cite{Uemura04,Kanigel04} while Yuan $et$ $al.$ observed a rather unusual behavior with negative curvature as shown in this paper.~\cite{Yuan05} In addition, the former two groups proposed unusually huge effective carrier mass.~\cite{Uemura04,Kanigel04}

On the other hand, several experiments have recently revealed that the CoO$_2$-layer thickness of this material varies sensitively depending on the Na content and the amount of H$_2$O molecules, and, furthermore, that the magnetic and SC properties are quite sensitive to the layer thickness.~\cite{Takada05,Sakurai05a,Sakurai05b,Lynn03,Sakurai04,Ihara04a,Ihara05a,Michioka06,Zheng06b,Sato06} Motivated by these findings, we previously studied how the variation of CoO$_2$-layer thickness affects the properties of this material.~\cite{Mochizuki06a,Mochizuki07} In these studies, we pointed out the following aspects; 
[1]: The slight layer-thickness variation induces drastic changes in the band structure and the Fermi surface (FS) topology. In particular, the FS is constructed from double $a_{1g}$-band cylinders around the $\Gamma$-point in the thick-layer systems, while in the thin-layer systems, it is constructed from one $a_{1g}$ cylinder around the $\Gamma$-point and six $e'_g$ pockets near the K-points. The former type of FS is referred to as FS1 while the latter is FS2.
[2]: The spin fluctuation is enhanced at $q\ne0$ in the thick-layer systems and at $q=0$ in the thin-layer systems.~\cite{Mochizuki07} The critical enhancement toward magnetic ordering (MO) is expected in the system with moderate thickness. In addition, $T$-dependence of the spin fluctuation also has strong layer-thickness dependence. The discrepancies of the NMR, NQR and $\mu$SR results are explained by this layer-thickness dependence.~\cite{Mochizuki07} [3]: Both singlet extended $s$-wave pairing~\cite{Kuroki06} and triplet $p$-wave pairing are possible in this material depending on the CoO$_2$-layer thickness or the FS geometry. The former is favored in the thick-layer systems with FS1, and the latter in the thin-layer systems with FS2 as shown before. The experimental phase diagram~\cite{Takada05,Sakurai05a,Sakurai05b} containing successive SC, MO and another SC phases is explained by this FS deformation.

In this letter, we discuss the thermodynamic properties of the SC states in Na$_x$CoO$_2 \cdot y$H$_2$O, and show that the scattered experimental results can be explained fairly well if we consider the two different SC states with different FS topologies. Certainly, the scattered $H_{c2}$ data are easily understood if we attribute the data exceeding the Pauli limit to the triplet $p$-wave pairing, and the other data to the singlet extended $s$-wave pairing. On the other hand, $C(T)$ and $\rho_s(T)$ are quantities sensitive not only to the SC gap structure but also to the electronic structures such as FS topology and density of states (DOS). Thus, in the following, we microscopically calculate these quantities (1) for the extended $s$-wave state on FS1, and (2) for the $p$-wave state on FS2. We show that the two different $C(T)$ data reported by Oeschler $et$ $al.$ and Jin $et$ $al.$ are reproduced for each pairing state. Two distinct behaviors of $\rho_s(T)$ data are also reproduced for each case. We also discuss the experimentally proposed large residual DOS and extremely huge effective mass of electrons. These results support our theoretical proposals about the two kinds of FSs in the present cobaltate system.

The calculations are performed in the same way as ref.~\citen{Nomura02}. We start with the multiorbital Hubbard model: $H_{\rm mo}=H_{\rm kin.}+H_{\rm int.}$. The first term $H_{\rm kin.}$ expresses electron hoppings between the Co $t_{2g}$ orbitals. In this term, we consider up to the third nearest neighbor hoppings, which are represented by twelve kinds of transfer integrals $t_1$, ..., $t_{12}$ and a crystal field (CF) parameter $\Delta$ for the trigonal CF from O ions. The detailed expression of $H_{\rm kin.}$ appears in ref.~\citen{Mochizuki07}. The second term $H_{\rm int.}$ is the Coulomb-interaction term, which consists of five contributions as $H_{\rm int.}=H_{U}+H_{U'}+H_{J_{\rm H}}+H_{J'}+H_{V}$. Here, $H_{U}$ and $H_{U'}$ are the intra- and inter-orbital Coulomb repulsions, respectively, and $H_{J_{\rm H}}$ and $H_{J'}$ are the Hund's-rule coupling and the pair hopping, respectively. These interactions are expressed using Kanamori parameters, $U$, $U'$, $J_{\rm H}$ and $J'$, which satisfy the relations; $U'=U-2J_{\rm H}$ and $J_{\rm H}=J'$. The last term $H_{V}=V\sum_{i,j}n_in_j$ expresses the Coulomb repulsion between electrons on the adjacent Co sites.

\begin{figure}[tdp]
\includegraphics[scale=1.0]{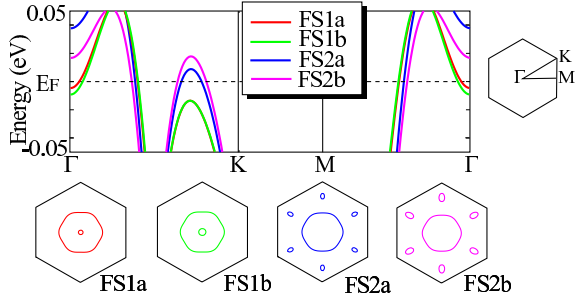}
\caption{Band dispersions near the Fermi level. Two types of the Fermi surfaces, FS1 (FS1a and FS1b) and FS2 (FS2a and FS2b), predicted in the previous study are reproduced.~\cite{Mochizuki06a,Mochizuki07}}
\label{Fig01}
\end{figure}
\begin{table}[b]
\begin{tabular}{c|ccccccc}
\hline
 & $t_1$& $t_2$& $t_3$& $t_4$& $t_5$& $t_6$& $t_7$ \\
\hline
 FS1a& 11.5& -21.1& 148.3& 43.8& -17.3& -10.3& 9.0  \\
 FS1b& 10.2& -20.8& 149.1& 43.3& -17.7& -11.5& 8.2  \\
 FS2a& 47.2& -25.4& 150.8& 49.0& -17.6& -14.5& 3.7  \\
 FS2b& 47.2& -25.4& 150.8& 49.0& -17.6& -14.5& 3.7  \\
\hline
\end{tabular}
\begin{tabular}{cccccc}
\hline
  $t_8$& $t_9$& $t_{10}$& $t_{11}$& $t_{12}$& $\Delta$ \\
\hline
  -53.3& -31.2& -24.6& 6.0& -1.2& 35.0 \\
  -53.1& -29.4& -25.0& 6.1& -1.2& 35.0 \\
  -56.6& -44.7& -26.8& 9.5& -7.3& 30.0 \\
  -56.6& -44.7& -26.8& 9.5& -7.3& 60.0 \\
\hline
\end{tabular}
\caption{Transfer integrals and crystal-field parameters. The energy unit is meV.}
\label{tbl:params}
\end{table}
The values of $t_1$, ..., $t_{12}$ and $\Delta$ are determined so as to reproduce the band structures which predicted the two types of FSs, i.e., FS1 and FS2.~\cite{Mochizuki06a,Mochizuki07} The obtained values are listed in Table.~\ref{tbl:params} where the energy unit is meV. Figure~\ref{Fig01} shows band structures with FS1 (FS1a and FS1b) and those with FS2 (FS2a and FS2b) near the Fermi level. The inner $a_{1g}$ cylinder of FS1b (the $e'_g$ pocket of FS2b) is slightly larger than that of FS1a (FS2a). The whole Co $t_{2g}$ band structures are shown in refs.~\citen{Mochizuki06a} and \citen{Mochizuki07}. Hereafter, the energy unit is eV.

We derive the effective pairing interactions for both spin-singlet and spin-triplet channels ($\hat{\Gamma}^{\rm s}(q)$ and $\hat{\Gamma}^{\rm t}(q)$) by using the random-phase approximation (RPA). Then, we construct the Eliashberg equation.
The detailed expressions of $\hat{\Gamma}^{\rm s}(q)$, $\hat{\Gamma}^{\rm t}(q)$ and the Eliashberg equation are included in refs.~\citen{Mochizuki07} and \citen{Mochizuki04}.
\begin{figure}[tdp]
\includegraphics[scale=1.0]{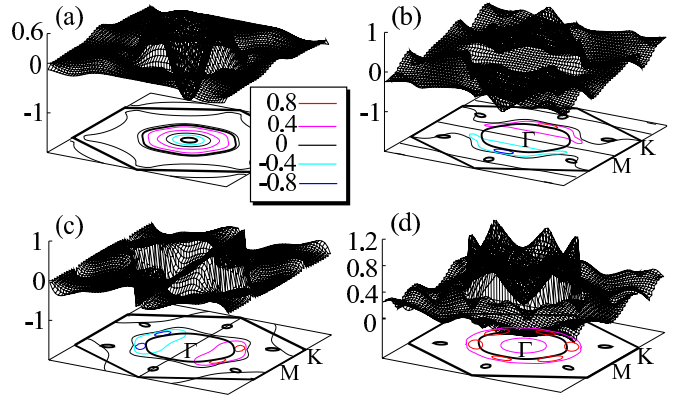}
\caption{Momentum dependence of the gap functions for (a) the extended $s$-wave state on FS1, $\phi_a^{es}({\bf k})$, (b) the $p_x$-wave state on FS2, $\phi_a^{px}({\bf k})$, (c) the $p_y$-wave state on FS2, $\phi_a^{py}({\bf k})$, and (d) the linear combination of the $p_x$-wave and the $p_y$-wave states, $\phi_a^{p}({\bf k})=\sqrt{\phi_a^{px}({\bf k})^2+\phi_a^{py}({\bf k})^2}$.}
\label{Fig02}
\end{figure}
By solving the Eliashberg equation, we obtain the momentum dependence of SC gap function $\phi_{mn}({\bf k})$. The calculations are numerically carried out by taking 128$\times$128 ${\bf k}$-meshes in the first Brillouin zone and 1024 Matsubara frequencies. We take $U$=0.50, $J_{\rm H}$=0.05, $V$=0.05, $k_{\rm B}T$=0.01 and the Co valence $s=$+3.4. Using the unitary matrix $\hat{U}({\bf k})$ which diagonalizes $H_{\rm kin.}$, we obtain the gap function with band basis ($\phi_a({\bf k})$) from those with orbital basis ($\phi_{mn}({\bf k})$). In Fig.~\ref{Fig02}, we display $\phi_a({\bf k})$ for (a) the extended $s$-wave state on FS1 ($\phi_a^{es}({\bf k})$), (b) the $p_x$-wave state on FS2 ($\phi_a^{px}({\bf k})$), and (c) the $p_y$-wave state on FS2 ($\phi_a^{py}({\bf k})$). 

As for the $p$-wave pairing, the $p_x$ and $p_y$ states are degenerate in the triangular lattice. In addition, there are three choices of the direction of the $d$ vector ($\hat{x}$, $\hat{y}$, $\hat{z}$), i.e., the $S_z$ component of the $S=1$ Cooper pair. Below $T_c$, a linear combination of these six states should be realized, namely, $\hat{\rm z}(p_x \pm ip_y)$, $\hat{\rm x}p_x \pm \hat{\rm y}p_y$ and $\hat{\rm x}p_y \pm \hat{\rm y}p_x$. All of these gaps are represented as $\phi_a^{p}({\bf k})=\sqrt{\phi_a^{px}({\bf k})^2+\phi_a^{py}({\bf k})^2}$, which is shown in Fig.~\ref{Fig02} (d).

In the following, we assume that the ${\bf k}$- and $T$-dependence of the SC
gap function below $T_c$ is described by
$\Delta_{{\bf k}a}(T)=\phi_a({\bf k}) \Delta(T)$.
Substituting this into the standard BCS gap equation,
\begin{equation}
\Delta_{{\bf k}a}(T)= - \frac{1}{N} \sum_{{\bf k'},a'} 
V_{{\bf k}a,{\bf k'}a'} \frac{\tanh [\frac{E_{{\bf k'}a'}}{2k_{\rm B}T}] }{E_{{\bf k'}a'}} 
\Delta_{{\bf k'}a'}(T),
\label{eq:BCSeq}
\end{equation}
with $V_{{\bf k}a,{\bf k'}a'}=-V\phi_a({\bf k}) \phi_{a'}^*({\bf k'})$ 
and $E_{{\bf k}a}=\sqrt{\xi_{{\bf k}a}^2+|\Delta_{{\bf k}a}(T)|^2}$,
we obtain,
\begin{equation}
1= -\frac{V}{N} \sum_{{\bf k},a} |\phi_a({\bf k})|^2
\frac{\tanh [\frac{E_{{\bf k}a}}{2k_{\rm B}T}] }{2E_{{\bf k}a}}.
\label{eq:BCSeq2}
\end{equation}
Considering that $\Delta(T_c)$=0, we calculate the temperature evolution of $\Delta(T)$ below $T_c$ by solving eq.~(\ref{eq:BCSeq2}). Along with $T_c$ of the actual material, we take $k_{\rm B}T_c$=0.45 meV ($T_c$$\sim$ 4.5 K) throughout our calculations. The entropy $S(T)$, $C(T)$ and $\rho_s(T)$ are calculated by the formula, 
$S(T)=-2\sum_{{\bf k},a} \{ [1-f(E_{{\bf k}a})] \log [1-f(E_{{\bf k}a})]
+f(E_{{\bf k}a}) \log f(E_{{\bf k}a}) \}$,
$C(T)=T \frac{\partial S(T)}{\partial T}$,
and
$\rho_s(T) \propto \sum_{{\bf k},a} |{\bf k}|^2 (
 \frac{\partial f(E_{{\bf k}a})}{\partial E_{{\bf k}a}}
-\frac{\partial f(\xi_{{\bf k}a})}{\partial \xi_{{\bf k}a}})$.

\begin{figure}[tdp]
\includegraphics[scale=1.0]{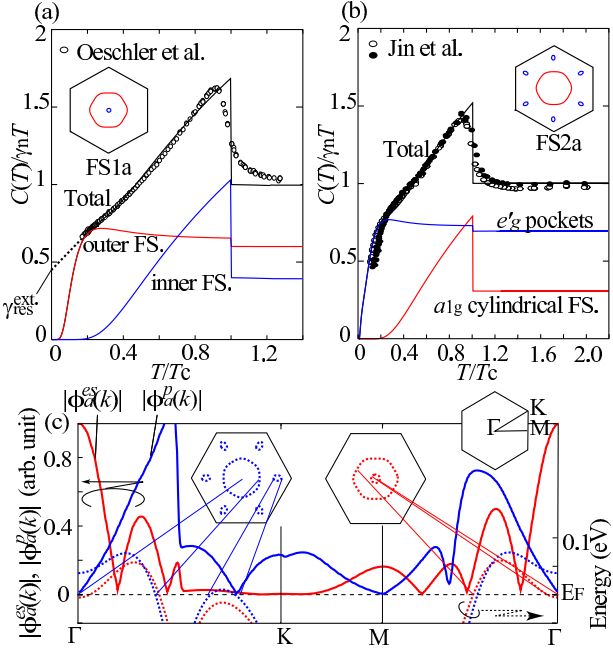}
\caption{Calculated specific heat $C(T)/\gamma_n T$ for (a) the extended $s$-wave state on FS1a and (b) the $p$-wave state on FS2a. The experimental data from refs.~\citen{Oeschler05} and \citen{Jin05} are also depicted respectively in each figure. A crude extrapolation gives a superficialy large $\gamma_{\rm res}^{\rm ext}$ as shown in (a) by the dashed line. (c) Momentum dependence of the gap functions of both pairing states, $\phi_a^{es}({\bf k})$ (red line) and $\phi_a^{p}({\bf k})$ (blue line), together with the band dispersions near the Fermi level for both FS1a and FS2a cases (red and blue dashed lines, respectively).}
\label{Fig03}
\end{figure}
Figures~\ref{Fig03} (a) and ~\ref{Fig03}(b) show the calculated Sommerfeld coefficient $\gamma(T)$=$C(T)/T$ normalized by the normal-state $\gamma$ ($\gamma_n$) for (a) the extended $s$-wave state on FS1a and (b) the $p$-wave state on FS2a, together with the experimental data from Oeschler $et$ $al.$ (ref.~\citen{Oeschler05}) and Jin $et$ $al.$ (ref.~\citen{Jin05}), respectively. Both figures show good coincidece between the experiment and the calculation. 
In the following, we discuss the results in detail.

\noindent
[1]: In Fig.~\ref{Fig03}(a), we show the contributions from the two parts of FS1. Apparently, the inner FS dominantly contributes to the jump at $T_c$ ($\Delta\gamma$) for the FS1 case. This is because the extended $s$-wave state on FS1 has a large gap amplitude on the inner FS. This can be seen in Fig.~\ref{Fig03}(c), in which $\phi_a^{es}({\bf k})$ is depicted together with the band dispersion for the FS1a case near the Fermi level. This figure also shows that the gap on the outer FS of FS1 is markedly small. Similarly, for the $p$-wave state on FS2, the cylindrical $a_{1g}$ FS dominantly contributes to $\Delta\gamma$ as shown in Fig.~\ref{Fig03}(b). This is due to a large gap amplitude on the $a_{1g}$ FS as shown in Fig.~\ref{Fig03}(c). In the present $p$-wave state, the gaps on the $e'_g$ pockets are markedly small. Note that this is in sharp contrast to the previous proposals of the $p$-wave or $f$-wave state with large gap amplitudes on the $e'_g$ pockets.~\cite{Mochizuki04,Kuroki04,Yanase04,Mochizuki05}

\noindent
[2]: For the $p$-wave state on FS2, the gap almost vanishes at certain points on the $e'_g$ pockets, which form line nodes. Consequnce of such a line-nodal gap is seen in the $T^2$-dependence of $C(T)$ ($\gamma(T)$=$C/T \propto T$) at low temperatures in Fig.~\ref{Fig03}(b). 

\noindent
[3]: Oeschler $et$ $al.$ reported $H_{c2}(T\rightarrow0)$ of $\sim$8 T, which is comparable to, and does not exceeds the Pauli limit of $\sim$8.3 T for $T_c$$\sim$4.5 K.~\cite{Oeschler05} This is in agreement with the spin-singlet pairing expected in their sample. On the other hand, Jin $et$ $al.$ reported $H_{c2}(T\rightarrow0)$ of 17.1 T, which apparently exceeds the Pauli limit.~\cite{Jin05} This is consistent with the expected spin-triplet pairing in their sample. 

\noindent
[4]: The $C(T)$ data of Jin $et$ $al.$ shows a deviation from linearity with steeper decrease below $T$$\sim0.8$ K ($T/T_c$=0.17 and $T_c$=4.7 K). Our result well reproduces this behavior as shown in Fig.~\ref{Fig03} (b). In contrast, Oeschler $et$ $al.$ did not observe such a decrease down to 0.85 K ($T/T_c$=0.19 and $T_c$=4.52 K). However, as shown in Fig.~\ref{Fig03} (a), if the temperature is further decreased, the steep decrease should be observed also for the extended $s$-wave case.

\noindent
[5]: Oeschler $et$ $al.$ claimed a large residual $\gamma$ ($\gamma_{\rm res}$) of 6.67 mJ mol$^{-1}$K$^{-2}$ at $T\rightarrow0$ by extrapolation. The estimated SC volume fraction $(\gamma_n-\gamma_{\rm res})/\gamma_n$ is small to be 59 $\%$ where $\gamma_n$=16.1 mJ mol$^{-1}$K$^{-2}$. In fact, a number of measurements have reported similar or larger $\gamma_{\rm res}$ so far.~\cite{Takada05} Our results suggest two reasons why the experimentally estimated $\gamma_{\rm res}$ is so large. One is possible existence of non-SC sections of the FS. Both extended $s$-wave and $p$-wave SC gaps have markedly small amplitude on the outer FS of FS1 and on the $e'_g$ pockets of FS2, respectively. These tiny gaps should be easily destroyed by impurities and/or defects. The resulting non-SC sections of the FS should generate large $\gamma_{\rm res}$. In fact, this possibility was also proposed by Takada $et$ $al$.~\cite{Takada05} A large residual DOS reported from NQR $^{59}(1/T_1T)$ measurements~\cite{Ishida03} may also be attributed to such non-SC FSs. Another reason is the crude extrapolation which overlooks the steep decrease at very low temperatures. This should cause the overestimate of $\gamma_{\rm res}$ as indicated by the dashed line in Fig.~\ref{Fig03}(a).

\begin{figure}[tdp]
\includegraphics[scale=1.0]{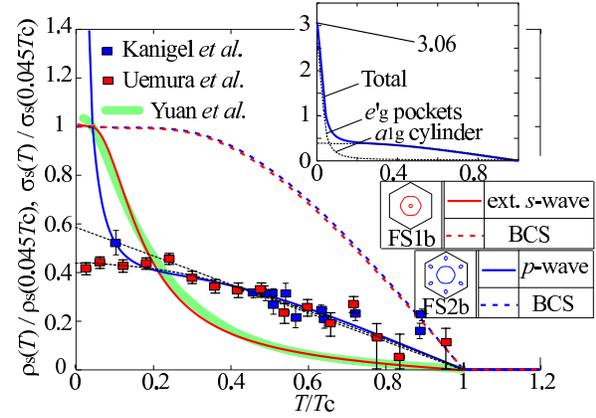}
\caption{Calculated superfluid density $\rho_s(T)/\rho_s(0.045T_c)$ and experimental $\sigma_s(T)$ data from Uemura $et$ $al.$~\cite{Uemura04} and Kanigel $et$ $al.$~\cite{Kanigel04}, and $\lambda(T)^{-2}$ data from Yuan $et$ $al$~\cite{Yuan05}. Black dashed lines are crude extrapolations of the data to $T\rightarrow0$. Inset: Calculated $\rho_s(T)/\rho_s(0.045T_c)$ ($\sigma_s(T)/\sigma_s(0.045T_c)$) for the $p$-wave state on FS2b. The contribution from the $a_{1g}$ FS and that from the $e'_g$ pockets are shown separately.}
\label{Fig04}
\end{figure}
Next, we show the calculated $\rho_s(T)$ in Fig.~\ref{Fig04} for (a) the extended $s$-wave state on FS1b, and (b) the $p$-wave state on FS2b, which are normalized by the value at $T$=0.045$T_c$. In addition, the results for the uniform BCS gap are shown by dashed lines for comparison. Both the extended $s$-wave and the $p$-wave cases exhibit unusual behaviors, which are completely different from the BCS case. Below $T_c$, both two start to increase rather gradually, and even at low temperatures, both exhibit increasing behaviors in contrast to the saturating behaviors of the BCS cases. We compare our calculations with the experimental data of $\mu$SR rate $\sigma_s(T)$ reported by Uemura $et$ $al.$ (ref.~\citen{Uemura04}) and Kanigel $et$ $al.$ (ref.~\citen{Kanigel04}), and the magnetic-penetration-depth ($\lambda(T)^{-2}$) data measured by Yuan $et$ $al.$ utilizing a tunnel diode based self-inductive technique (ref.~\citen{Yuan05}). Here, $\sigma_s(T)$ and $\lambda(T)$ are related to $\rho_s(T)$ as $\sigma_s(T) \propto \lambda(T)^{-2}=4\pi \rho_s(T) e^{2}/m^{\ast}c^{2}$ where $m^{\ast}$ is the effective mass of the carriers. We also depict their data with proper normalization in Fig.~\ref{Fig04}.

For the extended $s$-wave case, the calculated $\rho_s(T)$ (blue line) is convex downward, and shows quite good coincidence with the experimental data from Yuan $et$ $al.$ The $H_{c2}$ value measured in this experiment is less than the Pauli limit, which is consistent with the present spin-singlet pairing.
For the $p$-wave case, the calculated $\rho_s(T)$ (red line) shows an steeper increase starting from a lower temperature than the extended $s$-wave case and shows good agreement with the expreimental data from Kanigel $et$ $al.$ and Uemura $et$ $al$. However, the data from Uemura $et$ $al.$ slightly deviates from the calculation, which shows no steep increase at low temperatures. In fact, the steep increase is originated from the weakly gapped $e'_g$ pockets (See the inset. Contributions from each part of FS2 are depicted separately). Such small gaps would be easily destroyed by impurities and/or defects, resulting in the ${\it residual}$ normal-state carrier density $\rho_n$. This should cause the suppression of the increasing behavior of $\rho_s(T)$.~\cite{Note01}

Both Kanigel $et$ $al.$ and Uemura $et$ $al.$ estimated $\sigma_s(T)$ at $T\rightarrow0$ by extrapolation. They assumed that thus obtained $\sigma_s^{\rm ext}(0)$ corresponds to $\sigma_s(0)$ for the clean limit ($\sigma_s^{\rm clean}(0)$) where all the normal-state carriers above $T_c$ contribute to the superconductivity at $T=$0 (namely, $\rho_s(T=0)$=$\rho_n(T>T_c)$). Using this $\sigma_s^{\rm clean}(0)$ value, the effective carrier mass $m^{\ast}$ was estimated to be about 75 times the bare electron mass $m_e$ by Kanigel $et$ $al.$ and about 100 times by Uemura $et$ $al.$ These values are unusualy huge and comparable to the effective mass in the heavy-electron superconductors. However, the extrapolation leads to underestimate of $\sigma_s^{\rm clean}(0)$ when we overlook the subsequent steep increase at low temperatures or the possible large residual $\rho_n(0)$ arising from the non-SC $e'_g$ pockets due to the sample inhomogeneity. As a result, their extrapolations give only 13-20 $\%$ of the actual value of $\sigma_s^{\rm clean}(0)$. (Note that in Fig.~\ref{Fig04}, the crude extrapolations give $\sigma_s^{\rm ext}(0)/\sigma_s(0.045T_c)\sim$0.6 for Kanigel $et$ $al.$ and $\sim$0.4 for Uemura $et$ $al.$ wheareas the calculation gives $\sigma_s^{\rm clean}(0)/\sigma_s(0.045T_c)$=3.06.) These underestimates should cause approximately 5 and 7.5 times overestimates of $m^{\ast}$, respectively. Thus, the estimated effective mass should be $m^{\ast}\sim$15 $m_e$. This value is reasonable, and is in good agreement with that estimated from the angle-resolved photoemission spectroscopy.~\cite{ShimojimaPC}

To summarize, we have studied the $T$-dependence of $C(T)$ and $\rho_s(T)$ in the two SC states of Na$_x$CoO$_2 \cdot y$H$_2$O predicted in our previous calculations; the extended $s$-wave state on FS1 and the $p$-wave state on FS2. We have shown that the two different $C(T)$ data as well as the two distinct $\rho_s(T)$ data are reproduced for the two cases. This advocates our proposal of possible two different SC states in this material depending on the CoO$_2$-layer thickness or the FS geometry. In addition, there are weak SC sections of the FS for both cases; the outer FS of FS1 and the $e'_g$ pockets of FS2, respectively. These cause sharp decrease (increase) of $C(T)$ ($\rho_s(T)$) at very low temperatures, which makes the extrapolation of the data to $T\rightarrow0$ difficult, and generates large residual $\gamma$ and $\rho_n$ when these small gaps are destroyed by impurities and/or defects. The reported unusually large $\gamma_{\rm res}$ and the extremely huge $m^*$ are understandable from these facts. We lastly note that it is quite possible that the two superconducting states can coexist in the same sample, especially in large single crystals, due to the ununiformity of the CoO$_2$-layer thickness or the ununiform water distribution, as suggested from a recent impurity-effect measurement.~\cite{YJChenCD05}

We thank H. Sakurai, Y. Yanase, T. Shimojima, M. B. Salamon, and P. Badica for discussions. This work is supported by a Grant-in-Aid for Scientific Research from MEXT and the RIKEN special Postdoctoral Reseacher program. H.Q.Y. is supported by the US Department of Energy under award No. DEFG02-91ER45439.


\begin{thebibliography}{999}
\bibitem{Yuan05}Details of the experiment and more results will appear in another publication.
\end{thebibliography}
\end{document}